\documentclass[journal=jacsat,manuscript=communication,layout=twocolumn,articletitle=true]{achemso}

\usepackage[version=3]{mhchem} 
\usepackage{graphicx}
\usepackage[T1]{fontenc}       
\usepackage{dblfloatfix}
\usepackage{gensymb}
\usepackage{grffile}
\usepackage{chngcntr}
\usepackage{fixltx2e}
\usepackage{chemformula}
\usepackage[T1]{fontenc} 
\usepackage{subfig}
\usepackage{epsfig}
\usepackage{braket}
\usepackage{psfrag}
\usepackage{subfig}
\usepackage{multirow}
\usepackage{xr}
\usepackage{mathtools}
\usepackage{amsmath}
\usepackage{amssymb}
\usepackage{commath}
\usepackage{amsfonts}
\usepackage[normalem]{ulem}
\usepackage{braket}
\usepackage{array}
\usepackage{tabularx}
\usepackage{stackengine}

\setkeys{acs}{articletitle = true}

\author{Lorenzo A. Mariano}
\affiliation[]
{Grenoble-INP, SIMaP, University of Grenoble-Alpes, CNRS, F-38042 Grenoble, France}
\author{Bess Vlaisavljevich}
\affiliation[Second University]
{Department of Chemistry, University of South Dakota, Vermillion, South Dakota 57069, USA}
\author{Roberta Poloni}
\email{roberta.poloni@grenoble-inp.fr}
\affiliation[]
{Grenoble-INP, SIMaP, University of Grenoble-Alpes, CNRS, F-38042 Grenoble, France}

\title[An \textsf{achemso} demo]
      {Improved Spin-State Energy Differences of Fe(II) molecular and crystalline complexes {\sl via} the Hubbard $U$-corrected Density}

\abbreviations{IR,NMR,UV}

\begin{document}
\maketitle

\begin{abstract}
  We recently showed that the DFT+U approach with a linear-response $U$ yields adiabatic energy differences biased towards high spin [Mariano et al. {\sl J. Chem. Theory Comput.} {\bf 2020}, 16, 6755-6762]. Such bias is removed here by employing a density-corrected DFT approach where the PBE functional is evaluated on the Hubbard $U$-corrected density. The adiabatic energy differences of six Fe(II) molecular complexes computed using this approach, named here PBE$[$U$]$, are in excellent agreement with coupled cluster-corrected CASPT2 values for both weak- and strong-field ligands resulting in a mean absolute error (MAE) of 0.44 eV, smaller than the recently proposed Hartree-Fock density-corrected DFT (1.22 eV) and any other tested functional, including the best performer TPSSh (0.49 eV). We take advantage of the computational efficiency of this approach and compute the adiabatic energy differences of five molecular crystals using PBE$[$U$]$ with periodic boundary conditions. The results show, again, an excellent agreement (MAE=0.07 eV) with experimentally-extracted values and a superior performance compared with the best performers TPSSh (MAE=0.08 eV) and M06-L (MAE=0.31 eV) computed on molecular fragments.
\end{abstract}

The accurate description of spin-state energetics of transition metal complexes represents a great challenge for electronic structure ab initio methods\cite{wilbraham2017multiconfiguration,domingo2010spin,radon2019benchmarking,cirera2012theoretical,pierloot2006relative,swart2008accurate,droghetti2012assessment,cirera2018benchmarking}. Yet, the accurate prediction of spin-state energy differences are of critical importance for the understanding of spin crossover phenomena relevant for example for spintronics, molecular elecronics and sensors \cite{kumar2017emerging,molnar2019molecular,C9SC02522G,D0DT01533D} and for the catalytic reactivity of biological systems \cite{Shaik2011}.
This challenge stems from the lack of error cancellation when computing energy differences, using approximate electronic structure methods, between spin states exhibiting different types and amounts of electronic correlations (dynamic and non-dynamical)\cite{Neese2020}. Because Hartree-Fock (HF) only treats exchange correlations, for example, it tends to stabilize high-spin states over low spin states due to the absence of dynamical correlation that would stabilize doubly occupied orbitals \cite{Swart2008,Rehier2012}. On the contrary, local and semilocal functionals within DFT tend to overstabilze low spin states \cite{swart2004validation,Fouqueau2005,mortensen2015spin,ioannidis2015towards,ganzenmuller2005comparison,Salomon2002} and by adding a fraction of exact exchange one can, in most cases \cite{radon2014revisiting}, reduce such overstabilization \cite{Pinter2017,Prokopiou2018}. Thus, global hybrids can provide a reasonable decription of spin-state energetics depending on the system of choice and the amount of exact exchange \cite{Rehier2012,Prokopiou2018}.\\
Song at al. showed that a HF density-corrected DFT approach can yield adiabatic energy differences in good agreement with diffusion Monte Carlo (DMC) calculations\cite{song2018benchmarks}. The DFT+U approach has also been investigated in this respect in a few studies and we refer the reader to the introduction of Ref.~\citenum{Mariano2020} for a recent summary on the topic. The present authors have shown and discussed the bias towards high spin states imposed by the Hubbard term in the total energy and how it can be mitigated by adopting values of $U$ smaller than the computed self-consistent value, $U$\textsubscript{sc} \cite{Mariano2020}. Despite the energetics being pathologically wrong for strong-field ligands, the electronic density exhibits a systematic {\sl improvement} with respect to local and semilocal functionals for both low spin and high spin states and for all systems when adopting a self-consistent $U$ \cite{cococcioni2005linear,zhao2016global,Mariano2020}.
\begin{figure}[ht]
  \centering
  \includegraphics[scale=0.40]{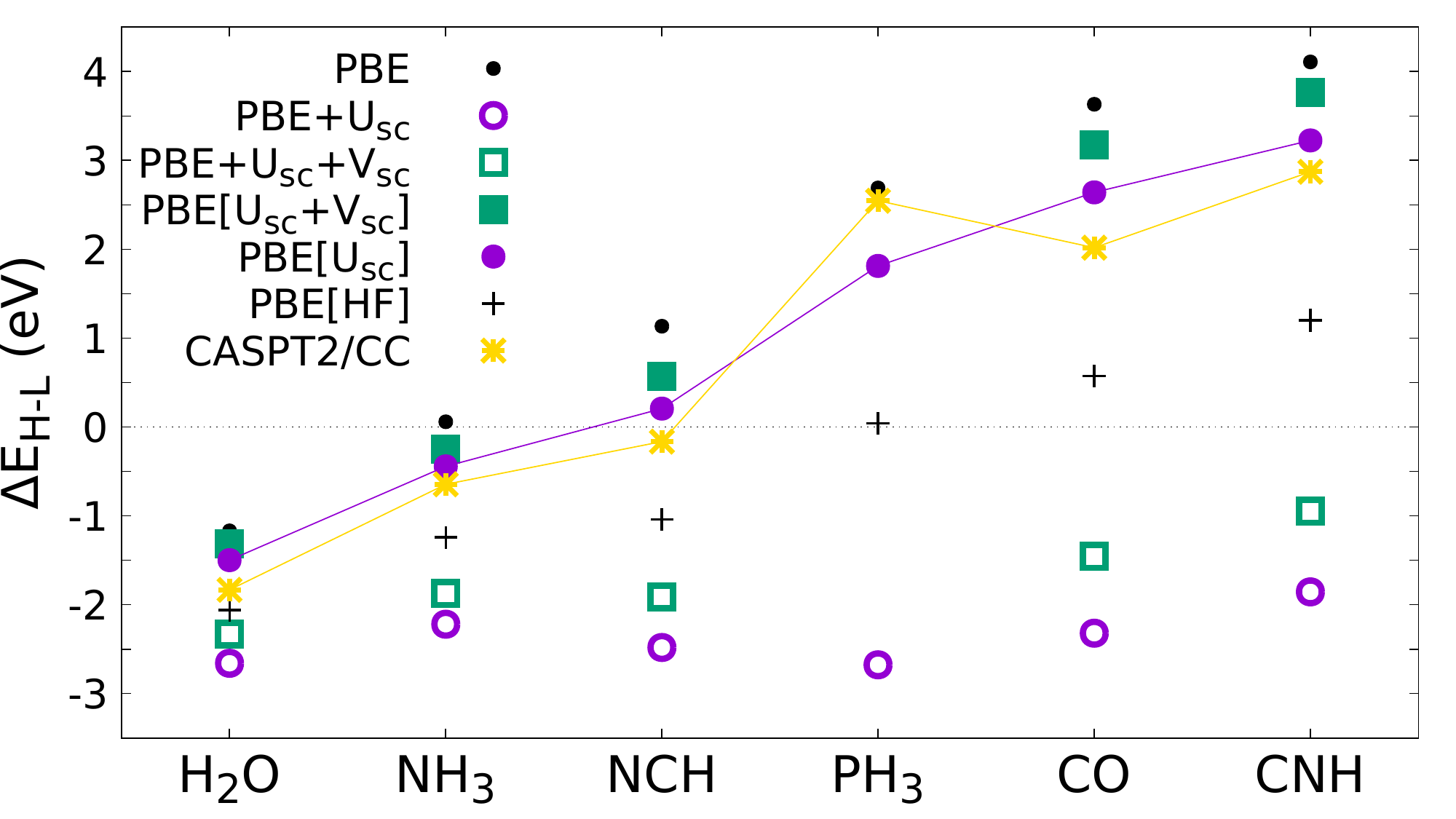}
 \includegraphics[scale=0.40]{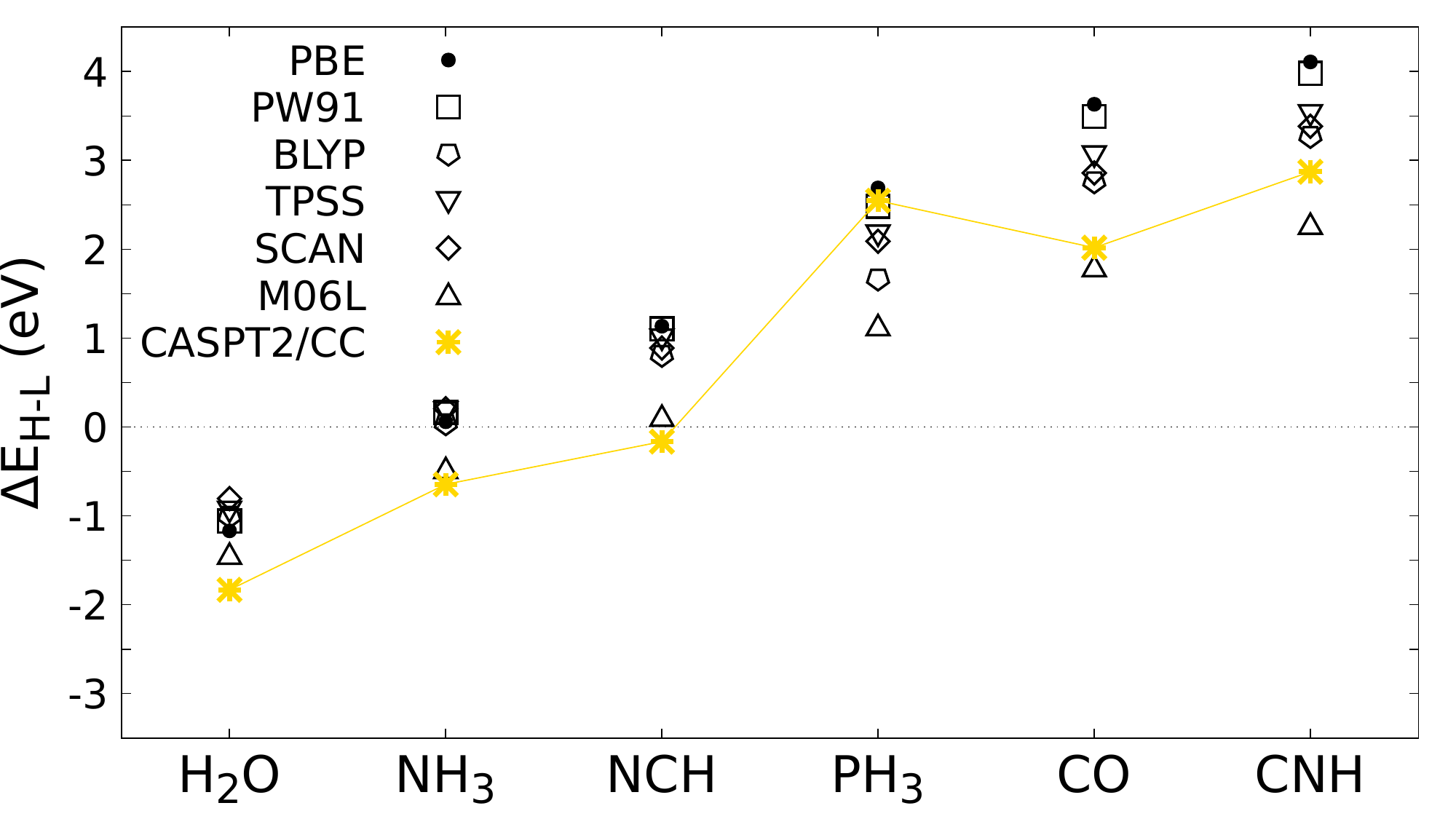}
 \includegraphics[scale=0.40]{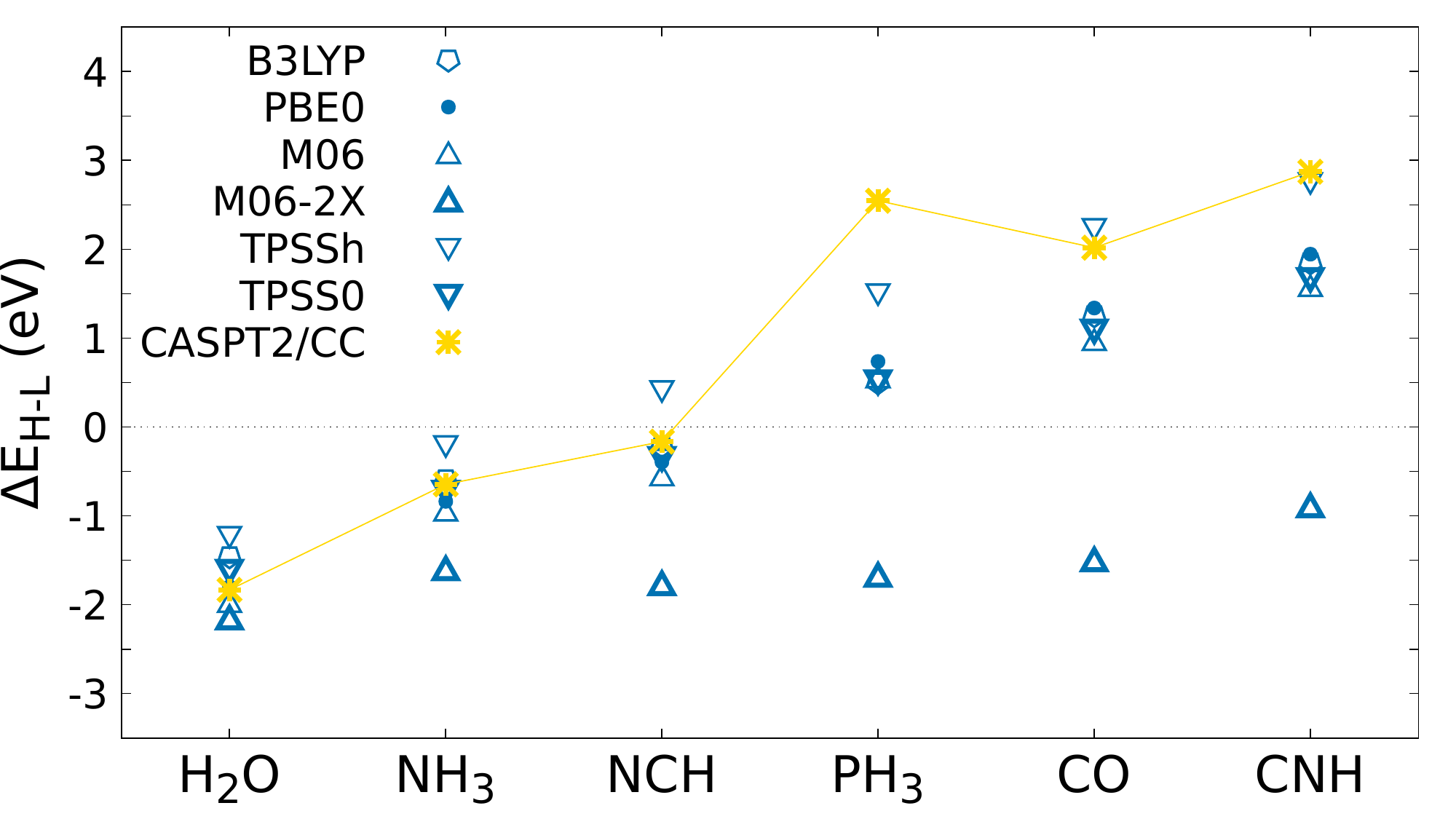}
\caption{Adiabatic energy differences, $\Delta E$\textsubscript{H-L}, computed using varying DFT approaches, together with the reference CASPT2/CC energies computed in this work. The values are also reported in Tab.~\ref{tab1} for clarity.}
\label{Fig1}
\end{figure}

In this work we merge the above ideas and adopt a new approach consisting of a Hubbard $U$ density-corrected DFT where the PBE functional is evaluated on the Hubbard $U$ density, using a linear-response $U$\cite{cococcioni2005linear} computed self-consistently\cite{Kul2006}. We show that this method allows one to obtain adiabatic energy differences for a series of six Fe(II) molecular complexes in excellent agreement with the chosen reference set. The molecular complexes include varying ligand field strengths from the weak H$_2$O ligand, whose reference  $\Delta E$\textsubscript{H-L} is -1.83 eV, to the strong CNH one, whose $\Delta E$\textsubscript{H-L} is 2.87 eV (see Tab.~\ref{tab1}), thus allowing for a better assessment of the validity of this approach. Although  the choice of the reference method is still matter of debate, we choose the coupled-cluster corrected CASPT2 approach proposed by Pierloot and coworkers \cite{pierloot2017spin,phung2018toward}. This approach reduces the overstabilization of high spin state by treating the 3s and 3p semicore electrons using CCSD(T) and can be used in principle in systems with non-negligible multiconfigurational character such as the strong-ligand field molecules studied here. Its accuracy has been recently further validated by Rado\'{n} by comparing with $\Delta E$\textsubscript{H-L} values extracted from experiments\cite{Rado2019}. We show that for Fe(II) complexes exhibiting a weak ligand strength, our result compare very well also with CCSD(T) \cite{domingo2010spin} and recent DLPNO-CCSD(T) results \cite{Neese2020}. Larger deviations are found with respect to DMC results ({\sl vide infra}).

\begin{table*}[ht!]
\centering
\renewcommand{\arraystretch}{1.3}
\begin{tabular}{|lcccccc|ccc|}
\hline \hline
                              &   \multicolumn{6}{c}{$\Delta E$\textsubscript{H-L}}  & \multicolumn{3}{c|}{MAE} \\ \hline
  DFT methods                 &   H$_2$O        & NH$_3$                  & NCH      & PH$_3$       &  CO        & CNH         &  weak-field & strong-field & total     \\ \hline
PBE$[$U$]$      &      -1.50      &           -0.44      &  0.21    & 1.81       &   2.64        &     3.23      & 0.30 & 0.57  & 0.44        \\
TPSSh                         &       -1.23     &        -0.21           &   0.41  &   1.51      &   2.24    &  2.76      & 0.53 & 0.45 & 0.49           \\
M06-L                          &       -1.44     &         -0.47            &  0.11   &   1.13     &    1.80    &   2.27     & 0.28 & 0.74 &   0.51        \\
PBE$[$U$]$ ({\footnotesize atomic proj.})              &      -1.33      &           -0.22      &  0.56    & 2.17       &   3.08        &     3.65      & 0.44 & 0.74  & 0.59        \\
PBE0                          &     -1.80       &        -0.84           &  -0.39    &  0.74       &  1.34     & 1.95      & 0.15 & 1.13 & 0.64           \\
TPSS0                         &       -1.61     &        -0.72            &  -0.34    &    0.52      &  1.09      &   1.67     & 0.01 & 1.38 & 0.70           \\
B3LYP                         &      -1.46      &       -0.59            & -0.21   &   0.50      &  1.25      &    1.85     & 0.16 & 1.28 &  0.72          \\ 
SCAN                          &       -0.81     &         0.21            &  0.89    &  2.09       &   2.86    &     3.38    & 0.97 & 0.60 &  0.79                \\
TPSS                          &       -0.94      &        0.18            &   1.00   &  2.17      &   3.06     &    3.52     & 0.96 & 0.69 &  0.82          \\
BLYP                          &      -1.00       &        0.04            &   0.81   &    1.67     &  3.06         &   3.52        & 0.83 & 0.85 & 0.84                 \\
M06                           &       -1.98     &        -0.95           &   -0.56  &     0.55     &  0.97     &    1.58     & 0.29 & 1.44 &    0.87       \\
PW91                          &      -1.06      &        0.16            &    1.10  &     2.48     &   3.49     &   3.98     & 0.94 & 0.88 & 0.91                \\
PBE                           &      -1.17      &           0.06          &   1.14     &     2.69        &   3.63        &     4.11       & 0.89 & 1.00 & 0.94                \\
PBE[HF]                       &      -2.06      &         -1.24           &    -1.04     &    0.04         &      0.58     &  1.20          & 0.57 & 1.87 &      1.22     \\
M06-2X                        &       -2.16     &        -1.61            &  -1.77   &   -1.68      & -1.51     &  -0.90      & 0.97  & 3.84 &  2.41       \\
\hline \hline
 Wavefunction methods                           &               &                         &            &             &            &                \\
\cline{1-7}\cline{1-7}
CASPT2/CC                     &     -1.83$^a$                    &       -0.64$^a$          &    -0.16$^a$            & 2.54$^a$       &   2.02$^a$      &     2.87$^a$ & \\ \cline{1-7}
 \multirow{4}{*}{CASPT2}   &  -1.99$^a$        &     -0.85$^a$         &     -0.27$^a$   &     2.31$^a$       &  1.78$^a$     &     2.66$^a$    \\
                            & -2.15\cite{wilbraham2017multiconfiguration}       &     -1.27\cite{wilbraham2017multiconfiguration}        &     -0.81 \cite{wilbraham2017multiconfiguration}    &         &  2.07\cite{wilbraham2017multiconfiguration}   &     2.78 \cite{wilbraham2017multiconfiguration}    \\ 
                             &  -1.88 \cite{domingo2010spin}        &     -0.98  \cite{domingo2010spin}         &    -0.32 \cite{domingo2010spin}     &     2.41\cite{domingo2010spin}       &  2.07 \cite{domingo2010spin}   &           \\
                              &   -2.02\cite{pierloot2006relative}      &  -0.88\cite{pierloot2006relative}            &        &            &       &        \\ \cline{1-7}
\multirow{2}{*}{CCSD(T)}                       &    -1.45  \cite{domingo2010spin}  &      -0.66  \cite{domingo2010spin} &   -0.19   \cite{domingo2010spin}&    1.51  \cite{domingo2010spin}  &   1.25 \cite{domingo2010spin}     &          \\
                                &     &      &    -0.09\cite{LawsonDaku2012}&     &     &          \\ \cline{1-7}
DLPNO-CCSD(T1)                &   -1.44 \cite{Flser2020} &  -0.49 \cite{Flser2020}       &  -0.38  \cite{Flser2020}  &           &         &         \\ \cline{1-7}
\multirow{3}{*}{DMC}                           &   -1.78  \cite{song2018benchmarks} &  -1.23  \cite{song2018benchmarks}   &  -1.17  \cite{song2018benchmarks}&            &   0.59 \cite{song2018benchmarks}  &         \\ 
                                       &   -2.60 \cite{Droghetti2012} &  -1.55 \cite{Droghetti2012}   & -1.37 \cite{Droghetti2012} &            &    &         \\ 
                                       &  &  & -0.85,-0.95 \cite{Fuma2016} &           &   &      \\ 
\cline{1-7} 
\end{tabular}
\caption{$\Delta E$\textsubscript{H-L} (in eV) computed using different DFT approaches (upper left table). These are reported in order of descreasing total MAE computed with respect to the CASPT2/CC reference (see text). The MAE computed separately for weak-, strong-ligand and for the whole set (total) are also reported (upper right table). The $\Delta E$\textsubscript{H-L} computed using varying wavefunction methods and taken from the literature are also reported (lower table). $^a$ refers to the CASPT2/CC values computed in this work. }
\label{tab1}
\end{table*}

We then apply this approach to compute the spin crossover energies of seven compounds, either crystalline or molecular, for which the adiabatic energy differences have been extracted from experiments and we find again very good agreement. In light of this accuracy, this approach can be adopted to study molecular crystals very efficiently with any DFT code including a DFT+U implementation thus avoiding the use of hybrid functionals. \\
We recall that the DFT+U total energy can be written as:
\begin{equation}
 E\textsubscript{DFT+U}[\rho(r)]=E\textsubscript{DFT}[\rho(r)]+E\textsubscript{U}[{n}]
\label{eq:u2}
\end{equation}{}
In the above formula the term E\textsubscript{DFT} represents the unperturbed DFT energy functional and the E\textsubscript{U} is the Hubbard term containing the Hubbard correction for the electronic repulsion within a given subshell and a double-counting term that removes the interactions that are already counted within the DFT term via mean-field. For a clear review we refer the reader to Ref.~\citenum{Himme2014}. The E\textsubscript{U} depends on the density through the occupation numbers ${n}$ computed from the projection of the occupied Kohn-Sham eigenfunctions onto a localized basis set. For projection numbers close to 1/2 the summation term that enters E\textsubscript{U} and that mutliplies $U$ is the largest.  In our recent work we showed that the DFT+U energy yields a systematic bias towards high spin due to the E$_U$ term being systematically larger for low spin states thus resulting in a destabilization of the latter with respect to the former. This bias increases as a function of the ligand field strength: for stronger field ligands the more covalent bonding between Fe and the ligand yields more fractional occupations thus resulting in larger penalizing summation terms\cite{Mariano2020}. While this penalizing term is necessary to recover the localization of electrons and stabilize the insulating phase in Mott physics, here it results in a systematic unphysical overstabilization of high spin which further increases for molecular complexes with larger covalent character, such as the CO and CNH strong field ligands.\\
Cococcioni and coworkers implemented an extended Hubbard model in DFT through the inclusion an inter-site effective interaction $V$ within the Hubbard energy. Such a generalized scheme, named DFT+U+V \cite{2010CocoDFTUV}, aims at an improved treatment of electronic correlations. The new Hubbard potential includes two terms of opposite sign: the first on-site term is attractive for Kohn-Sham states that exhibit a localized character (the standard on-site $U$ term) whereas the second inter-site term stabilizes hybridized states. Thus, a competition between these two opposing behaviors should allow for a more balanced description of electronic correlations and thus improved structural and electronic properties \cite{2010CocoDFTUV,Marz2020}.
For the six Fe(II) molecular complexes computed here, i.e. [Fe(H$_2$O)$_6$]$^{+2}$, [Fe(NH$_3$)$_6$]$^{+2}$, [Fe(NCH)$_6$]$^{+2}$, [Fe(PH$_3$)$_6$]$^{+2}$, [Fe(CNH)$_6$]$^{+2}$ and [Fe(CO)$_6$]$^{+2}$, the geometries optimized using the TPSSh functional are taken from Ref.~\citenum{Mariano2020} and used for all calculations, i.e. DFT, CCSD(T) and CASPT2. All DFT calculations, except for the DFT+U, were performed using ORCA\cite{orca,orca_update}. The DFT+U and DFT+U+V calculations were performed using Quantum ESPRESSO\cite{quantumespresso,QE2017} by adopting a linear-response approach\cite{cococcioni2005linear} for the self-consistent calculation of $U$\cite{Kul2006}, i.e. $U$\textsubscript{sc},  and $U$+$V$\cite{timrov2018hubbard}, i.e. $U$\textsubscript{sc}+$V$\textsubscript{sc}, respectively. We stress that in what follows, DFT+U or DFT+U+V always refer to self-consistent calculations, unless otherwise specified (e.g. in the results discussed later in Fig.~\ref{Fig4}). See SI for more details. Unlike our recent work where a few geometrical optimizations were performed upon calculation of the linear-response $U$ to yield a structurally consistent $U$\cite{Mariano2020}, here $U$ and $U$+$V$ are computed on the TPSSH geometries for LS and HS separately. Because in what follows we report errors computed as deviations from the reference values, we intend to avoid including effects arising from different geometries. The effect of the employed geometry on the $\Delta E$\textsubscript{H-L}=$E$\textsubscript{HS}-$E$\textsubscript{LS} has also been investigated and will be discussed below. The projections for the Hubbard term are performed using orthonormalized atomic wavefunctions. This yields $\Delta E$\textsubscript{H-L} systematically smaller than those computed with non-orthogonal non-normalized atomic projectors (as those we employed previously \cite{Mariano2020}), as shown in Tab.~\ref{tab1} and Tab.~S2.
\par
Extended multi-state CASPT2 calculations were performed employing BAGEL\cite{bagel, bagel2} using an active space of 10 electrons in 12 orbitals, (10e,12o). This includes the 3d electrons of Fe(II), the two ligand-$e_g$ molecular orbitals plus the Fe 4d double-shell \cite{pierloot2006relative,wilbraham2017multiconfiguration}, and their corresponding electrons. Density fitting was used for all calculations by employing the fitting basis set cc-pV5Z-JKFIT and no symmetry constraints were imposed. The CASPT2 calculations used for the reference set were performed without any ionisation potential-electron affinity (IPEA) shift to the zeroth-order Hamiltonian. Because of the well established slow convergence of the CASPT2 energy with respect to basis set size, we perform the extrapolation of the spin-state energies to the complete basis set (CBS) limit. This is done separately for CASSCF and the CASPT2 energies, by adopting the three-point extrapolation method described in Refs.~\citenum{Feller1992,Feller1993,Helgaker1997}. The cc-pVTZ-DK, cc-pVQZ-DK, and cc-pV5Z-DK basis sets were used for this and the corresponding CASPT2 $\Delta E$\textsubscript{H-L} are reported in Tab.~S1.\\
For [Fe(H$_2$O)$_6$]$^{+2}$ we were unable to converge a (10e,12o) active space where the two ligand-$e_g$ orbitals remained in the active space for LS. The Fe 3s orbital consistently rotated into the active space replacing one of the ligand $e_g$ orbitals. This implies that for this particular molecule using the smaller active space should not impact the HS-LS energy splitting significantly, as reported by Gagliardi and co-workers\cite{wilbraham2017multiconfiguration} who reported CASPT2 values of -2.14 eV with (6e,10o) and -2.15 eV with (10e,12o). Differences of the order of 0.1 eV are reported in Ref.~\citenum{pierloot2006relative}. Thus, for water the extrapolation to CBS is performed using a (6e,10o) active space. See computational methods in the SI for more details.\\
The Fe semicore 3s3p correlation energy is computed using CCSD(T) by including and freezing the 3s3p electrons \cite{phung2018toward}, using ORCA. This correction is then added to the CASPT2 energy difference to yield the CASPT2/CC energy difference. The aug-cc-pwCVTZ-DK and cc-pVDZ basis sets were used for Fe and the ligand atoms, respectively (see details in SI). Extrapolation to the CBS limit is not required here as demonstrated by Pierloot and coworkers\cite{phung2018toward}.\\
\begin{figure}[ht]
  \centering
 \includegraphics[scale=0.43]{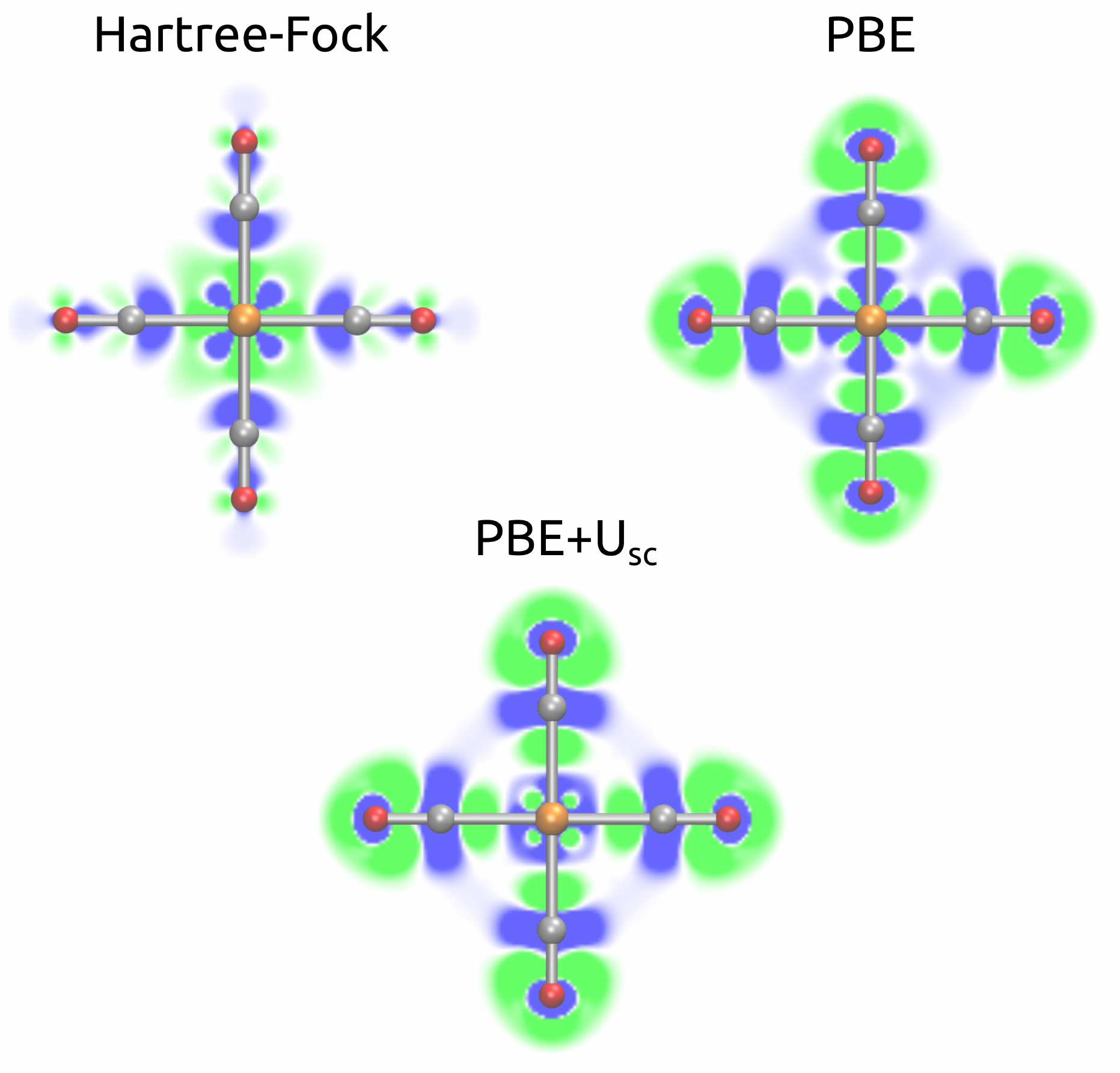}
 \caption{Density difference plot, $\delta \rho_{\text{x}}(r)$, for  [Fe(CO)$_6$]$^{+2}$ between x=PBE, PBE+U\textsubscript{sc}, and Hartree-Fock and the relaxed CASPT2 density; green and blue correspond to positive and negatives values, respectively. The plot shows values between -0.005 $e$/bohr$^3$ and 0.005 $e$/bohr$^3$.}
\label{Fig2}
\end{figure}
The adiabatic energy differences, $\Delta E$\textsubscript{H-L}, computed using several choices of DFT functionals including the DFT+U and DFT+U+V approaches are shown in Fig.~\ref{Fig1} together with the CASPT2/CC set. The PBE+U energies show an almost constant behavior throughout the molecular series\cite{Mariano2020} due to the penalizing Hubbard term being larger for LS and for strong-field ligands~\cite{Mariano2020}. A minor improvement of DFT+U+V as compared to DFT+U is found, possibly due to the values of $V$ being too low to correct for the bias towards HS (see Tab.~S3). For [Fe(PH$_3$)$_6$]$^{+2}$, we were unable to converge the DFT+U+V calculations for HS and thus the corresponding $\Delta E$\textsubscript{H-L} is omitted.\\
Despite yielding erroneous spin-state energetics for the molecular series reported here, the DFT+U with a linear-response $U$ approach systematically improves the electronic density, regardless of the spin state, with a reduction of the energy {\sl bowing} as a function of fractional occupations which is a manifestation of self-interaction error \cite{cococcioni2005linear,zhao2016global,Mariano2020}.
Song at al. \cite{song2018benchmarks} discuss the case of spin gaps in Fe(II) octahedrally-coordinated complexes in terms of calculations affected by large errors in the density: the error that arises from the approximation of the exchange-correlation functional is comparable or smaller than the error introduced by the use of the approximate density \cite{Sim2018}. In this respect, the density-corrected DFT approach, discussed in detail in Refs.~\citenum{2017Annual,Vuckovic2019}, consists in employing approximate density functionals on a density different than the self-consistent one and possibly closer to the {\sl exact} one. This approach implemented using the Hartree-Fock density has been shown to improve over the self-consistent DFT results the description of many properties such as reaction barriers\cite{Janesko2008, Verma2012}, weak intermolecular forces \cite{Gordon1972}, bond energies \cite{Kim2018} and the binding properties of anions \cite{Kim2011,Kim2014}.
The authors of Ref.~\citenum{song2018benchmarks} showed systematically improved results for spin-state splittings of fours Fe(II) molecular complexes computed using the DFT$[$HF$]$ approach. Our working hypothesis is that the Hubbard $U$-corrected density should yield more accurate results compared to a HF density, since the latter only includes exchange correlations while neglecting dynamic and non-dynamical correlations. We employ a density-corrected DFT by adopting a standard semilocal functional, such as PBE, evaluated on the Hubbard $U$ density. By doing so, we remove the energy bias introduced by the  E\textsubscript{U} term discussed above while keeping an improved electronic density.
Practically, we perform self-consistent DFT+U calculations by computing $U$ using linear-response theory and then remove the  E\textsubscript{U} term from the total energy to compute $\Delta E$\textsubscript{H-L}. This is not, strictly speaking, a non-self consistent, density-corrected DFT calculation as the kinetic term is computed using the DFT+U orbitals. However, our assumption is that the kinetic energy computed using DFT orbitals matches closely the DFT+U case so that this approach can be seen as a non-self consistent density-corrected DFT method. This assumption is motivated by a recent study employing Kohn-Sham inversion schemes to show that the Kohn-Sham kinetic energy and the Hartree-Fock one are negligibly different when computed on the same HF density \cite{Nam2020}.\\
The results of the Hubbard $U$\textsubscript{sc}-corrected density employed using a PBE functional are shown in Fig.~\ref{Fig1} and are named PBE$[$U$]$ henceforth. In the same figure we also show the PBE$[$HF$]$ results, i.e. using the Hartree-Fock density. We stress that the PBE+U and the PBE values are slightly different compared to those reported in our previous work \cite{Mariano2020} because of the different geometries employed and the atomic basis used for the projections. 
The PBE$[$U$]$ results are in excellent agreement with the reference CASPT2/CC set and provide a systematically improved description of $\Delta E$\textsubscript{H-L} compared with PBE$[$HF$]$. Specifically, PBE$[$HF$]$ yields a reasonable prediction for weak-field ligands but it performs poorly for strong field ligands. Our computed values of PBE$[$HF$]$ energies are similar to those reported in Ref.~\citenum{song2018benchmarks} on the same molecular complexes (i.e. [Fe(H$_2$O)$_6$]$^{+2}$, [Fe(NCH)$_6$]$^{+2}$, [Fe(NH$_3$)$_6$]$^{+2}$ and [Fe(CO)$_6$]$^{+2}$), however, our conclusion on the accuracy of DFT$[$HF$]$ is somehow different owing to the difference in the corresponding reference values. In particular, the DMC values in Ref.~\citenum{song2018benchmarks} are systematically lower compared to CASPT2/CC values and the largest deviation is found for the CO and NCH. See Tab.~\ref{tab1} for the whole list of $\Delta E$\textsubscript{H-L} computed either here or in previous studies using wavefunction methods. We note the reasonably good agreement between our CASPT2/CC reference values and published CCSD(T)\cite{domingo2010spin,LawsonDaku2012} and DLPNO-CCSD(T)~\cite{Neese2020} values for weak-field molecules (see Tab.~\ref{tab1}).\\
The performance of varying DFT functionals for the calculation of adiabatic energy differences has been reported in the literature by several authors \cite{pierloot2006relative,Pierloot2008,mortensen2015spin,swart2008accurate,ioannidis2015towards,wilbraham2017multiconfiguration,Ioannidis2017,song2018benchmarks,Pinter2017,Rado2019,Flser2020,Wilbraham2018,Alipour2020,Prokopiou2018,cirera2012theoretical,cirera2018benchmarking,Cirera2014,Vela2020,swart2004validation,Cirera2020,Droghetti2012}. Thus, we do compute the $\Delta E$\textsubscript{H-L} using a few DFT functionals in order to establish a comparison for the performance of DFT$[$U$]$, however, we refer the reader to these articles for a more detailed discussion.  GGA functionals overstabilize the LS state, although BLYP does so to a lesser extent compared to PBE and PW91. Excellent results have been reported in the past \cite{swart2008accurate,Pierloot2008} using the optimized OPTX exchange proposed by Handy and Cohen \cite{HANDY2001}. By adopting global hybrids with increasing admixtures of exact exchange HS is systematically stabilized. PBE0 and B3LYP with a 25\% and 20\% admixture of exact exchange added respectively to PBE and BLYP functionals \cite{mortensen2015spin} show an overcorrection and overall overstabilize HS.
Among the meta-GGAs, M06-L performs well in comparison to other meta-GGA functionals as already observed in previous studies \cite{ioannidis2015towards,Ioannidis2017,Cirera2020,Flser2020}. Among the studied functionals, the smallest error is found for the meta-hybrid TPSSh (15\% of exact exchange) in agreement with several recent studies \cite{cirera2012theoretical,Cirera2014,cirera2018benchmarking}. Climbing up the DFT Jacob's ladder other functionals like double-hybrid \cite{Wilbraham2018,Alipour2020} and range-separated hybrid functionals \cite{Prokopiou2018} have been tested.  Kronik et al. \cite{Prokopiou2018} employed optimally tuned range-separated hybrid functionals to study [Fe(H$_2$O)$_6$]$^{+2}$, [Fe(NCH)$_6$]$^{+2}$, [Fe(NH$_3$)$_6$]$^{+2}$ and [Fe(bpy)$_3$]$^{+2}$ and found good agreement with the available CCSD(T) and CASPT2 reference values. In Refs.~\citenum{Wilbraham2018} and \citenum{Alipour2020} the PBE0-DH-based double-hybrid and SOS0-PBESCAN0-2(a) double-hybrid were employed, respectively, to study [Fe(H$_2$O)$_6$]$^{+2}$, [Fe(NCH)$_6$]$^{+2}$, [Fe(NH$_3$)$_6$]$^{+2}$ and [Fe(CO)$_6$]$^{+2}$. The authors reported a good agreement when comparing with the DMC reference values of Ref.~\citenum{song2018benchmarks}. If these values are compared with our CASPT2/CC results, a systematic underestimation of $\Delta E$\textsubscript{H-L} is observed.
In Tab.~\ref{tab1} we report the $\Delta E$\textsubscript{H-L} of each method tested here and the MAE computed with respect to CASPT2/CC.  Among the employed approaches, PBE$[$U$]$ is the best performer with the lowest total MAE of 0.44 eV. It follows TPSSh with a MAE of 0.49 eV and M06-L with MAE of 0.51. PBE$[$U$]$ with values computed using atomic projectors represents the forth best performer. For the weak-field molcules (i.e. H$_2$O, NH$_3$ and NCH), TPSS0, PBE0 and B3LYP yield the best agreement with the reference with MAEs of 0.01 eV, 0.15 eV and 0.16 eV, respectively. These are followed by M06-L (0.28 eV) and PBE$[$U$]$ (0.30 eV). We note that PBE$[$HF$]$ yields a MAE similar to TPPSh for these three molecules (0.57 eV). For strong-field ligands (i.e. PH$_3$, CO and CNH), the best performers are TPSSh (MAE=0.45 eV) and PBE$[$U$]$ (MAE=0.57 eV) followed by M06-L and PBE$[$U$]$ with atomic projectors, while PBE$[$HF$]$ is the second worst performer after M06-2X (MAE=1.87 eV).\\
\begin{figure*}[ht]
  \centering
 \includegraphics[scale=0.8]{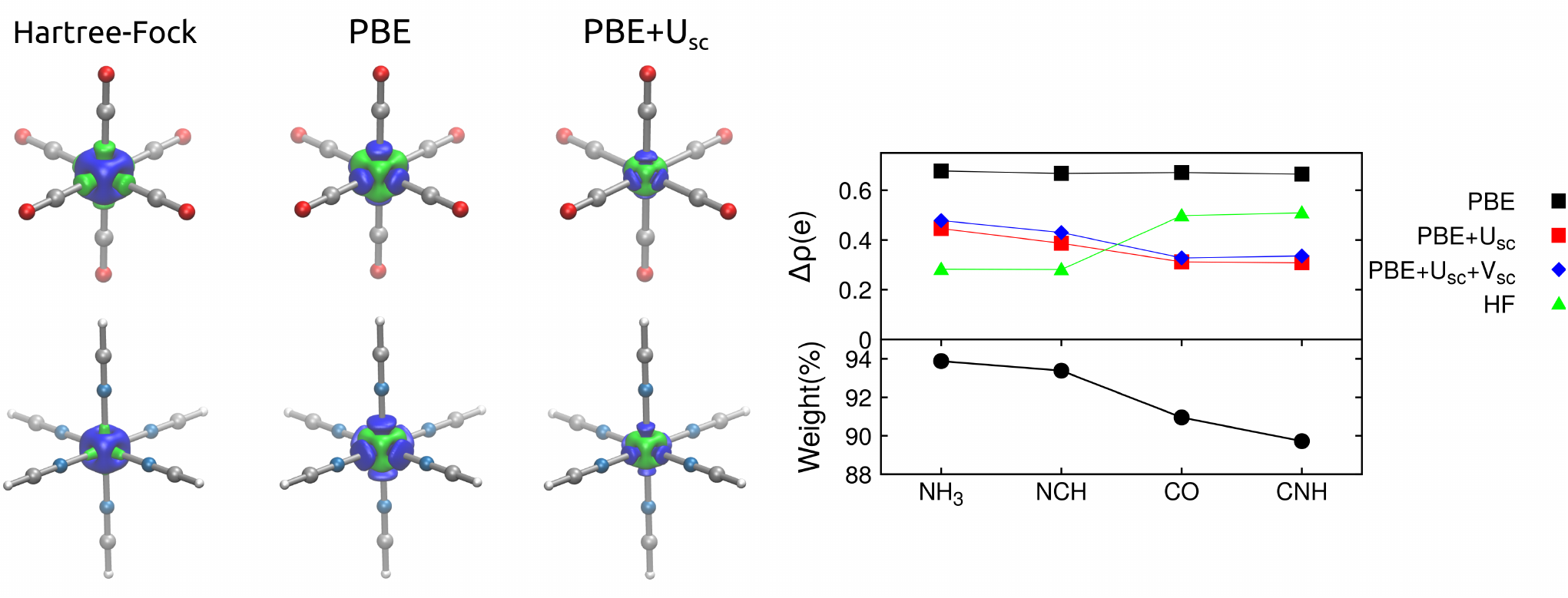}
 \caption{Left figure: charge density difference, $\rho^{Fe,Bader}_{\text{x}}(r)-\rho^{Fe,Bader}_{\text{CAS}}(r)$, plotted for CO and NCH ligands for x=HF, PBE, and PBE+U\textsubscript{sc}. Green and blue correspond to positive and negative isovalues of 0.004 $e$/bohr$^3$. HF and PBE yield an opposite behavior, with the former yielding a lower deviation from the reference for NCH as compared to CO. Right figure: error on the density estimated by computing $\Delta\rho_{\text{x}}$ (see text), i.e. by integrating the difference in charge density within the Bader region (upper panel). Weight (in \%) of the dominant electronic configuration within the CASSCF wavefunction (lower panel).}
\label{Fig3}
\end{figure*}
 The role of the geometry on the computed $\Delta E$\textsubscript{H-L} was also investigated. For each of the 11 functionals tested above we compute the $\Delta E$\textsubscript{H-L} using the geometries optimized with PBE+U, PBE, TPSSh, B3LYP and PBE0. We do this for [Fe(NH$_3$)$_6$]$^{2+}$ and  [Fe(CO)$_6$]$^{2+}$. The PBE+U geometry is optimized using a procedure that iteratively computes $U$ and then relaxes the geometry with this $U$ until convergence is achieved.  Overall, a non-negligible effect of the geometry on the $\Delta E$\textsubscript{H-L} is found (see Tabs.~S4 and S5). The $\Delta E$\textsubscript{H-L} change at most by 0.08 eV for [Fe(NH$_3$)$_6$]$^{2+}$, and 0.25 eV for [Fe(CO)$_6$]$^{2+}$, if the PBE+U geometry is excluded. When the PBE+U geometry is considered, the largest deviation is 0.18 eV and 0.78 eV for the weak- and strong-field ligand molecules, respectively. Specifically, regardless of the functional used to compute $\Delta E$\textsubscript{H-L}, the PBE+U geometry always yields the largest decrease in $\Delta E$\textsubscript{H-L}. This is consistent with the fact that the PBE+U geometry computed using a structurally consistent approach deviates the most from the TPSSh optimized geometry as shown in Tabs.~S6 and S7. Because the effect of $U$ is larger for LS \cite{Mariano2020} due to the larger values of E$_U$ as compared to HS, any other functional would destabilize LS more than HS. For stronger-field molecules this effect is more pronounced as confirmed by the larger increase in metal-ligand bond distances in the PBE+U LS geometry with respect to PBE (Tabs.~S6 and S7). \\
 To understand why the Hubbard $U$-corrected PBE density yields significantly improved results compared with the PBE density and the HF density, we compare densities from PBE, PBE+U, PBE+U+V (again computed with $U$\textsubscript{sc} and $V$\textsubscript{sc}), and Hartree-Fock density with that obtained from the relaxed CASPT2 spin-density matrix in BAGEL~\cite{Toru2016}. The relaxed spin-density matrix is obtained by adding orbital and configurational relaxation contributions due to dynamical correlation to the unrelaxed density matrix using the CASPT2 Lagrangian~\cite{Toru2016}.  We study the LS case of [Fe(NH$_3$)$_6$]$^{2+}$, [Fe(NCH)$_6$]$^{2+}$, [Fe(CO)$_6$]$^{2+}$ and [Fe(CNH)$_6$]$^{2+}$.
In Fig.~\ref{Fig2} we plot the difference $\delta \rho_{\text{x}}(r) = \rho_{\text{x}}(r) - \rho_{\text{CASPT2}}(r)$ between the electronic density obtained with x=[PBE, PBE+U, HF], and the CASPT2 relaxed density, for [Fe(CO)$_6$]$^{2+}$. The same qualitative result is obtained for NCH and the corresponding plots are reported in Fig.~S1. A limitation of this analysis is that large density differences are found in the spatial region near the ligand for PBE and PBE+U, while negligible ones are found for the HF density, consistent with a CASSCF active space mostly involving states associated with the Fe, and only marginally associated with the ligand, i.e. the two $e_g$ ligand states. Thus, the CASPT2-relaxed density resembles closely the Hartree-Fock one near the ligand, which is the reference method used to get the CASSCF wavefunction. Due to this limitation, in what follows we limit our considerations to the spatial region near the iron. When PBE density is used, the $\delta \rho_{\text{\tiny{PBE}}}(r)$ is negative within the spatial region associated with the $e_g$ orbitals and positive within for the $t_{2g}$ one indicating charge depletion and accumulation, respectively, for PBE compared with CASPT2. We note that CASPT2 calculations with the inclusion of bonding metal-ligand $e_g$ states and the 4d double shell have been shown to account for non-dynamical correlation \cite{PIERLOOT2003,pierloot2006relative,pierloot2001computational,Pierloot2011}. DFT does not account for non-dynamical correlation, however the self-interaction error arising from the implementation of approximate density-functionals yields an overdelocalization of the charge density along the chemical bonds (and less charge near the atom) and a more diffuse character of the electron cloud around the nuclei, as shown in Fig.~\ref{Fig2}, that can actually mimic these effects \cite{Polo2002,Graf2004,Harvey2006} sometimes called left-right and radial correlations, respectively. As shown in Fig.~\ref{Fig2}, and as discussed in the literature, these effects are exaggerated in PBE. The PBE+U\textsubscript{sc} density is qualitatively similar to the PBE density but with a smaller deviation from the reference one within the Fe region. We chose not plot the PBE+U+V density because it yields a plot visually identical to the PBE+U one. The Hartree-Fock density exhibits the opposite behavior near the Fe, i.e. charge density accumulates and depletes with the $e_g$ and $t_{2g}$ orbitals, respectively. This is consistent with the lack of explicit non-dynamical correlations and absence of self-interaction error. Thus, the effect of the Hubbard $U$ term on the density is qualitatively similar to the case found when increasing the exact exchange admixture in global hybrid functionals \cite{Pinter2017}.\\
 \begin{figure}[hb!]
   \centering
   \includegraphics[scale=0.43]{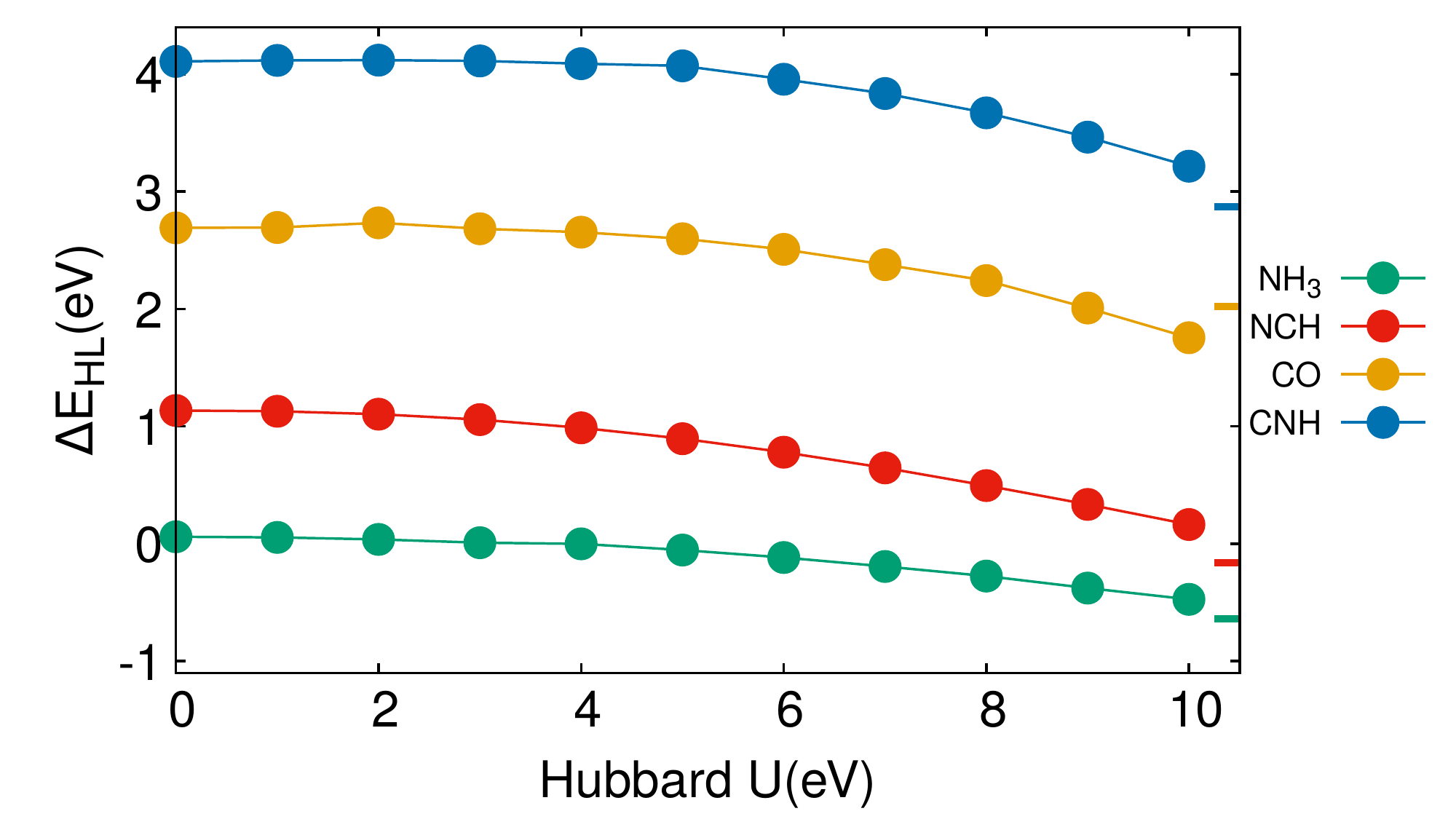}
   \centering
   \caption{Adiabatic energy differences, $\Delta E$\textsubscript{H-L}, computed with PBE$[$U$]$ as function of Hubbard $U$. On the right $y$-axis the CASPT2/CC reference values are shown by horizontal vertical dashes.}
	\label{Fig4}
\end{figure}
In what follows we attempt to quantify the error on the density by considering, again, only the region around the Fe. For each method, we extract the electronic charge density distribution around the iron centre, $\rho^{Fe}_{\text{x}}(r)$, by employing the Bader scheme \cite{Henkelman2006,Sanville2007,Tang2009,Yu2011}. We evaluate the error on the density, $\Delta\rho_{\text{x}}$, as a deviation from the reference, $\rho^{Fe}_{\text{CASPT2}}(r)$, within the Bader region (see SI for more details), as follows:
\begin{equation}
  \Delta\rho_{\text{x}} = \int \abs{\rho^{Fe,Bader}_{\text{x}}(r)-\rho^{Fe,Bader}_{\text{CAS}}(r)}dr
\label{eq:u1}
\end{equation}{}
The charge density difference, $\rho^{Fe,Bader}_{\text{x}}(r)-\rho^{Fe,Bader}_{\text{CAS}}(r)$, is plotted in Fig.~\ref{Fig3} for a weak and a strong-field ligand case, i.e. NCH and CO. For each molecule the opposite behavior of HF and PBE density is clearly visible, together with a reduced deviation from the reference density of PBE+U\textsubscript{sc}, in agreement with Fig.~\ref{Fig2} and the above considerations. The $\Delta\rho_{\text{x}}$ computed for the four molecules is plotted in the right panel of Fig.~\ref{Fig3}. This metric gives a constant error throughout the molecular series for PBE. The error associated with the HF density is smaller compared to PBE, and significantly smaller for weak-field molecules. This result is consistent with the reasonable prediction of $\Delta E$\textsubscript{H-L} found when employing PBE$[$HF$]$ for weak-field ligands and with the larger MAE of PBE for both weak- and strong-field ligands (see Tab.~\ref{tab1}).
\begin{table*}[ht!]
\centering
\renewcommand{\arraystretch}{1.3}
\begin{tabular}{|l|cc|ccccc|}
  \hline \hline
 \multirow{2}{*}{Complex}     &   \multicolumn{2}{c|}{$\Delta E$\textsubscript{H-L} / Periodic} & \multicolumn{5}{c|}{$\Delta E$\textsubscript{H-L} / Gas phase}   \\ \cline{2-8}
                              &   exp.          & PBE[U\textsubscript{sc}] & exp.                & PBE[U\textsubscript{sc}]   &   TPSSh         & M06-L         & PBE[HF]      \\ \hline
Fe(phen)$_2$(NCS)$_2$     &  0.155$^a$          &     -0.065           &  0.093$^a$          &  -0.117                    &   0.372$^a$     & -0.151$^a$   &  -0.887    \\
Fe(abpt)$_2$(NCS)$_2$     &  0.137$^a$          &      0.086           &  0.156$^a$          &  -0.032                    &   0.433$^a$     &  0.121$^a$   &  -1.006    \\
Fe(abpt)$_2$(NCSe)$_2$    &  0.150$^a$          &      0.159           &  0.184$^a$          &  -0.009                    &   0.491$^a$     &  0.115$^a$   & -0.950      \\
Fe[HB(pz)$_3$]$_2$        &  0.223$^a$          &      0.179           &  0.363$^a$          &  0.251                     &   0.722$^a$     &  0.428$^a$   &  -0.757     \\
FeL$_2$[BF$_4$]$_2$     &  0.198$^a$          &      0.196           &  0.208$^a$          &  0.191                     &   0.574$^a$     &  0.150$^a$   &  -0.671     \\ 
$[$Fe(tacn)$_2]^{2+}$         &                     &                      &  0.165$^c$          &  0.166                     &   0.443         &    0.171    &  -0.727       \\
$[$Fe(bpy)$_3]^{2+}$          &                     &                      &  0.434-0.744$^b$    &  0.466                     &   0.858         &    0.513    &  -0.626     \\
\hline \hline
MSE                           &                     &       -0.062        &                      &  -0.120                    &    0.305        &    -0.059    &  -1.055       \\
MAE                           &                     &        0.065        &                      &   0.121                    &    0.305        &     0.079    &  1.055         \\
\hline \hline
\end{tabular}
\caption{Adiabatic energy difference (eV), mean signed error (MSE) and mean absolute error (MAE) (in eV) computed with different DFT methods. The reference values are extracted from experimental (exp.) data. $^{a}$: Ref.~\citenum{Vela2020}; $^{b}$: Ref.~\citenum{LawsonDaku2005}; $^{c}$: Ref.~\citenum{Rado2019}. The reference value used to calculate MSE and MAE for [Fe(bpy)$_3]^{2+}$ is 0.589 eV.}
\label{tab2}
\end{table*}
The $\Delta\rho_{\text{x}}$ increase for molecules with increasing ligand-field strengths when x=HF and the opposite is found for x=[PBE+U, PBE+U+V]. This behavior is consistent with the DFT+U approach correcting the density more for strong-field ligand molecules, as shown and discussed previously~\cite{Mariano2020}. 
The trend along the four molecules correlates with trends in non-dynamical correlations. In agreement with previous studies \cite{domingo2010spin, pierloot2006relative, PIERLOOT2003}, we find that moving along the spectrochemical series non-dynamical correlation becomes more important. The configuration interaction weight of the dominant electronic configuration computed from the CASSCF calculation decreases from 94\% to 89\% going from NH$_3$ to CNH (see lower panel of Fig.~\ref{Fig3}). This analysis is in line with our results showing HF to perform better for molecular complexes with weak-field ligands and lower non-dynamical correlation. One would thus expect that HF density would overstabilize HS compared to LS more for strong field molecules, which is indeed the case here (see Fig.~\ref{Fig1} and Tab.~\ref{tab1}). 
PBE+U (with $U$\textsubscript{sc}) systematically improves the electronic density for both weak and strong-field ligand molecules thus yielding an improved description of the spin-state energetics throughout the spectrochemical series. This is further shown in Fig.~\ref{Fig4} where we report the $\Delta E$\textsubscript{H-L} computed using the PBE functional evaluated on the PBE+U density, for increasing values of $U$. We only show the results for four complexes for clarity. Higher values of Hubbard $U$ stabilize HS more compared to LS, as expected, and the deviation of $\Delta E$\textsubscript{H-L} from the reference value (shown as an horizontal line on the right y-axis) systematically decreases as $U$ increases.

The effect of $U$-corrected density on the spin energetics is qualitatively similar to the effect observed when adopting densities computed with increasing amounts of exact exchange \cite{Rado2014,song2018benchmarks,Pinter2017}. It must be noted, however, that the change in $\Delta E$\textsubscript{H-L} reported here is significantly larger than those computed with a density-corrected approach using hybrid functionals \cite{Rado2014,Pinter2017}.\\
To further test the validity of PBE$[$U$]$, we compute $\Delta E$\textsubscript{H-L} for a set of seven Fe(II) compounds for which the HS-LS energy differences have been extracted from experimental data. The first five compounds, Fe(phen)$_2$(NCS)$_2$\cite{Gal1990} (phen=1,10-phenanthroline), Fe(abpt)$_2$(NCS)$_2$, and Fe(abpt)$_2$(NCSe)$_2$ from Ref.~\citenum{MOLINER1999279} with abpt=4-amino-3,5-bis(pyridin-2-yl)-1,2,4-triazole, Fe[HB(pz)$_3$]$_2$\cite{Bou2009} (pz=pyrazolyl), and FeL$_2$[BF$_4$]$_2$\cite{Hal2001} (L=2,6-di(pyrazol-1-yl)pyridine), are molecular crystals for which Vela et al.~\cite{Vela2020} have extracted the experimental $\Delta E$\textsubscript{H-L} by removing the (computed) vibrational contribution from the the measured total enthalpy change. The other two are molecular complexes, [Fe(tacn)$_2$]$^{+2}$ (tacn= 1,4,7-triazacyclononane) and [Fe(bpy)$_3$]$^{+2}$ (bpy=2,2'-bipyridine). The $\Delta E$\textsubscript{H-L} of [Fe(tacn)$_2$]$^{+2}$ has been extracted by Rado\'{n} \cite{Rado2019} using an approach similar to Ref.~\citenum{Vela2020}. The spin gap of [Fe(bpy)$_3$]$^{+2}$ has been extracted by Casida et al. from the light-induced population of the high-spin state \cite{LawsonDaku2005}. For all these complexes we first adopt a molecular model to compute the $\Delta E$\textsubscript{H-L}. For the five molecular crystals, this is done by carving a structure from the fully optimized geometry using periodic boundary conditions, similarly to the procedure adopted in Ref.~\citenum{Vela2020}. The geometrical optimization is performed using Quantum Espresso using the PBE functional together with the semiempirical Grimme's D3 correction \cite{Grimme2010} combined with the Becke-Johnson (BJ) damping scheme \cite{Smith2016}. The Hubbard $U$\textsubscript{sc} is then computed on the optimized geometry, using periodic boundary conditons. More details are reported in the SI.
For [Fe(tacn)$_2$]$^{+2}$ and [Fe(bpy)$_3$]$^{+2}$ the structure is optimized using TPSSh with ORCA. 
The gas phase calculations of $\Delta E$\textsubscript{H-L} computed using PBE$[U]$, PBE$[$HF$]$, TPSSh, and M06-L are reported in Tab.~\ref{tab2} together with the experimentally-extracted reference value. For the five crystals, the TPSSh and M06-L results are taken from Ref.~\citenum{Vela2020}. For the rest of the calculations (i.e. PBE$[$HF$]$ and PBE$[$U$]$ on the seven molecules and TPSSh and M06-L on the last two) we add the D3 correction (similar to Vela et al.~\cite{Vela2020}) with the BJ damping scheme, except for M06-L for which this is not implemented. We note the use of four significant digits in Tab.~\ref{tab2}, compared to three in Tab.~\ref{tab1}: the choice in Tab.~\ref{tab1} was made for consistency with the approximation reported in the values taken from the literature. We choose however to add a significant figure in Tab.~\ref{tab2} because the reported values are closer.
PBE$[$U$]$ and M06-L are the best performers with a MAE of 0.12 eV and 0.08 eV, respectively. They both slightly underestimate the $\Delta E$\textsubscript{H-L} resulting in negative values of the mean signed error (MSE). TPSSh systematically overestimates the adiabatic energy differences with a MAE and MSE of 0.31 eV. PBE$[$HF$]$ yields the largest error with a MAE of 1.06 eV. Consistent with the study of the six molecular complexes reported above, PBE$[$HF$]$ systematically underestimate the $\Delta E$\textsubscript{H-L} for these intermediate-/strong-field molecules resulting in the wrong prediction of the ground state for the whole set under study. 
The five molecular crystals were also studied using a full periodic approach using PBE$[$U$]$ within the D3+BJ approximation for the dispersion forces. Compared to the gas phase calculations, the only difference is the molecular versus periodic model because the energy functional and the $U$ computed on the LS and HS periodic geometries ({\sl vide supra}) are the same. PBE$[$U$]$ with periodic boundary conditions represents the best performers with a MAE of 0.07 eV, i.e. slightly smaller compared to the same calculation performed on molecular fragments. This results confirms the good performance of PBE$[$U$]$ established above using ab initio data as reference, and it shows its potential for the effecient calculation of adiabatic energy differences in crystalline complexes.
\par
In conclusion, we show that the PBE$[$U$]$ approach consisting of adopting the PBE functional evaluated on the PBE+U density, with a self-consistent approach for the calculation of $U$, represents a reliable and computationally efficient method for the calculation of spin gaps of both molecular complexes and molecular crystals. We show that for the six Fe(II) molecular complexes ranging from weak- (H$_2$O) to strong-field ligands (CNH) the MAE associated with the PBE$[$U$]$ is the smallest among all the studied DFT approaches, including the TPSSh and M06-L functionals. The MAE is computed using the CASPT2/CC calculations as reference values. The performance of the PBE$[$U$]$ approach is further validated by the good agreement with CCSD(T) energy differences computed for weak-field molecules and reported in the literature.
The PBE$[$HF$]$ approach that uses the PBE functional on the HF density shows a  reasonable agreement with reference values for weak-field molecules but a poor  performance for strong-field molecules. The calculations performed on five periodic crystals and two additional molecules for which experimentally extracted values are available confirm all these findings.

\begin{acknowledgement}
This work was funded by the French National Agency for Research (ANR-15-CE06-0003-01). Calculations were performed using resources granted by GENCI under the CINES grant number A0020907211. The froggy platform of the CIMENT infrastructure was also employed for the calculations. B.V. was supported by the Department of Energy, Basic Energy Sciences (DE-SC0019463). Some of the computations supporting this project were performed on High Performance Computing systems at the University of South Dakota, funded by NSF Award 1626516. Also resources of the National Energy Research Scientific Computing Center (NERSC), a U.S. Department of Energy Office of Science User Facility operated under Contract No. DE-AC02-05CH11231, were used for this project.
\end{acknowledgement}

\bibliography{bibtex}
\end{document}